# Pressure Dependence of the Dynamic Crossover Temperatures in Protein and its Hydration Water

Xiang-qiang Chu[1], Antonio Faraone[2,3], Chansoo Kim[1], Emiliano Fratini[4], Piero Baglioni[4], Juscelino B. Leao[2], and Sow-Hsin Chen[1,*]

[1]Department of Nuclear Science and Engineering, Massachusetts Institute of Technology, Cambridge, MA 02139
[2] NIST Center for Neutron Research, Gaithersburg, MD 20899-8562
[3]Department of Material Science and Engineering, University of Maryland, College Park, MD 20742
[4]Department of Chemistry and CSGI, University of Florence, Sesto F.no, Florence, I-50019

Recently we have shown experimental evidence for a fragile-to-strong dynamic crossover (FSC) phenomenon in hydration water around a globular protein (lysozyme) at ambient pressure. In this letter we show that when applying pressure to the protein-water system, the FSC crossover temperatures in hydration water of lysozyme tracks the similar Widom line emanating from the existence of a liquid-liquid critical point in a 1-D confined water (in MCM-41-S). The mean squared displacements (MSD) of hydrogen atoms in lysozyme and in its hydration water show a sudden change of slopes at the same characteristic temperature, which decreases with an increasing pressure. These results taken together give support of the idea that the dynamic crossover (or so-called glass transition) of the protein is a function of both temperature and pressure, following the FSC of its hydration water.

For a hydrated protein at ambient pressure, it is known that below a characteristic temperature $T_D$ = 220 K, it transforms into a glassy state and loses its conformational flexibility, showing no appreciable biological functions [1,2]. At and above $T_D$, this flexibility is restored and the protein is able to sample more conformational sub-states, thus becomes biologically active. This "dynamic crossover" in protein is traditionally detected by observing the changing of the slope of the mean squared displacement (MSD) of hydrogen atoms vs. T plot [1-4], and believed to be triggered by their strong coupling with their hydration water, which shows a similar dynamic crossover at approximately the same temperature [5-9].

In our previous experiments at ambient pressure, we measured the average translational α-relaxation time $\langle\tau_T\rangle$ of the hydration water molecules by qausi-elastic neutron scattering (QENS) and found that this dynamic crossover in hydration water occurs at a universal temperature $T_L$ = 225 ± 5 K in three bio-molecules--lysozyme protein[10], B-DNA[11] and RNA[12], and can be described as a fragile-to-strong dynamic crossover (FSC) [13]. Thus we have shown that $T_D \approx T_L$ at ambient pressure. Furthermore, previous experiments on confined water in MCM-41-S porous silica material [14,15] have shown that an increased applied pressure will shift the FSC temperature to a lower value. We now show that this is also true for the interfacial water on the surfaces of protein. We further show that while a well-defined FSC phenomenon is observed for the applied pressure up to 1500 bar, when exceeding this pressure, the FSC phenomenon disappears. We thus identify a Widom line [16] in the T-P plane with an end point for the case of protein hydration water which is nearly identical to that of the confined water in MCM-41-S. This implies the existence of liquid-liquid critical point in both the 1-D and 2-D confined water.

We use hen egg white lysozyme (L7651, three times crystallized, dialysed and lyophilized) obtained from Fluka, without further purification. The protein powder was extensively lyophilized to remove any water left. The dried protein powder was then hydrated isopiestically at 5°C by exposing it to water vapor in a closed chamber until h = 0.3 is reached (i.e. 0.3 g $H_2O$ per g dry lysozyme). The hydration level was determined by thermo-gravimetric analysis and also confirmed by directly measuring the weight of absorbed water. This hydration level was chosen to have almost a monolayer of water covering the protein surface [17]. A second sample was then prepared using $D_2O$ in order to subtract out the incoherent signal from hydrogen atoms of the protein. Both hydrated samples had the same water or heavy water/dry protein molar ratio. Differential scanning calorimetry (DSC) analysis was performed in order to detect the absence of any feature that could be associated with the presence of bulk-like water.

In this experiment, we measured both $H_2O$ hydrated sample and $D_2O$ hydrated sample, and take their difference to obtain the signal contributed solely from hydration water [18,19]. Using the High-Flux Backscattering Spectrometer (HFBS) in NIST Center for Neutron Research (NCNR), we were able to determine the temperature and pressure dependences of $\langle\tau_T\rangle$ for the hydration water. For the chosen experimental setup, the spectrometer has an energy resolution of 0.8 $\mu$eV and a dynamic range of $\pm 11$ $\mu$eV [20], in order to be able to extract the broad range of relaxation times from 1 ps to 10 ns over the temperature and pressure range. A specially designed thick-wall aluminum pressure cell, which enables neutrons to penetrate easily, was used in the high-pressure system. Helium gas, the pressure-supplying medium, fills the whole sample cell, and applies pressure to both hydrated samples. At each pressure, the experiment was done with a series of temperatures, covering both the fragile and the strong regimes of the relaxation times from measured spectra. Altogether, over two-year period, 2000 spectra were collected, spanning six pressures: ambient, 400, 800, 1200, 1500 and 1600 bars.

These pressures are below the pressure range where the cold denaturation of proteins is observed (3600 bars) [21].

We use the Relaxing Cage Model (RCM) [22] to extract $\langle \tau_T \rangle$ of the hydration water from the difference QENS data. The RCM describes the translational dynamics of water at supercooled temperature in terms of the product of two functions:

$$F_H(Q,t) = F^S(Q,t) \exp[-(t/\tau_T(Q))^\beta] \quad (1)$$
$$\tau_T(Q) \cong \tau_0 (0.5 Q)^{-\gamma} \quad (2)$$

where the first factor, $F^S(Q,t)$, represents the short-time vibrational dynamics of the water molecule in the cage. This short time intermediate scattering function (ISF) is calculated in the Gaussian approximation using the known density of states of hydrogen atoms in water. The second factor, the α-relaxation term, contains the stretch exponent $\beta$, and the Q-dependent translational relaxation time $\tau_T(Q)$, which is a strong function of temperature. $\tau_T(Q)$ is further specified by two phenomenological parameters $\tau_0$ and $\gamma$, the pre-factor and the exponent controlling the power-law Q-dependence of $\tau_T(Q)$ respectively. The average translational relaxation time, which is a Q-independent quantity we use in this paper, is defined as $\langle \tau_T \rangle = \tau_0 \Gamma(1/\beta)/\beta$, where $\Gamma(x)$ is the gamma function. The temperature dependent $\langle \tau_T \rangle$ is calculated from two fitted parameters, $\tau_0$, and $\beta$, by analyzing a group of quasi-elastic peaks at different Q values simultaneously using three Q-independent parameters, $\tau_0$, $\beta$, and $\gamma$. For each pressure measurement, we chose seven spectra from data taken at HFBS at each temperature. The small Q values of these spectra make it possible to neglect the contribution from the rotational motion of water molecule in ISF [22].

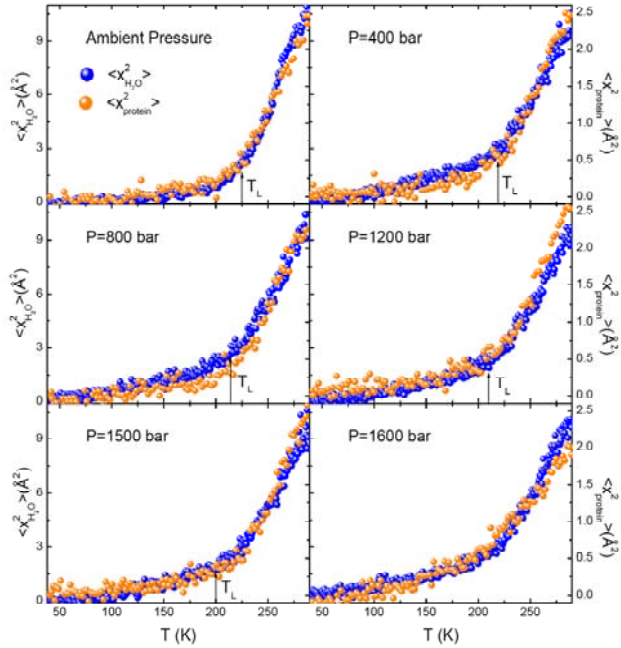

Fig. 1. The MSD, $\langle x^2 \rangle$, of hydrogen atoms in lysozyme and in its hydration water, as a function of temperature, under different pressures. Blue circles indicate the data processed from the difference between the $H_2O$ and $D_2O$ hydrated samples, which gives MSD of the H-atoms in the hydration water, following the scales on the left. Orange circles represent the data processed from the $D_2O$ hydrated sample, which gives MSD of H-atoms in the protein, following the scales on the right. The arrow signs indicate the crossover temperatures taken from Fig. 3.

To obtain the MSD $\langle x^2 \rangle$ of hydrogen atoms, we measure elastic scattering in the temperature range from 40 K to 290 K, covering completely the supposed crossover temperature $T_L$. Since the system is in a stationary metastable state at temperature below and above $T_L$, we make measurements by heating and cooling respectively at a heating/cooling rate of 0.75 K/min and observe exactly the same results. We calculate the MSD from Debye-Waller factor, $S_H(Q,\omega = 0) = \exp[-Q^2 \langle x^2 \rangle]$, by linearly fitting the logarithm of $S_H(Q,\omega = 0)$ with $Q^2$, where $S_H(Q,\omega = 0)$ can be easily calculated by taking the ratio of the temperature dependent elastic scattering intensity $I_{el}(Q,T,\omega = 0)$ and its low temperature limit,

$S_H(Q,\omega = 0) = I_{el}(Q,T,\omega = 0) / I_{el}(Q,T = 0,\omega = 0)$.

The calculated MSD of the hydrogen atoms in the hydration water molecule, $\langle x^2_{H2O} \rangle$, and that of the lysozyme molecule $\langle x^2_{protein} \rangle$, as a function of temperature, are shown in Fig. 1. The observational time interval is about 2 ns, corresponding to the energy resolution of 0.8 μeV. Comparing the results measured at 6 different pressures, we see clearly that the temperature dependence of the MSDs of lysozyme and its hydration water all follows the same trend, especially after we rescale them into the same amplitudes by multiplying the MSD of protein by a factor of 4.2. Each MSD shows a linear behavior at lower temperatures, but above a certain temperature, the slope abruptly increases. We can estimate a crossover temperature $T_L$ from the turning point of these curves. Note that the dynamic crossover temperature of the hydration water ($T_L$) and the dynamic crossover temperature of the lysozyme molecule ($T_D$) are visually the same, implying that the dynamic crossover in the protein is triggered by the dynamic crossover in its hydration water. Since the crossover temperature $T_L$ in the hydration water is pressure dependent, which will be shown clearly in figure 3, this leads to a conclusion that the dynamic crossover temperature $T_D$ in proteins follows the same pressure dependence.

A much sharper definition of this dynamic crossover temperature $T_L$ can be obtained from RCM analysis of the difference spectra of the two hydrated samples. Fig. 2 shows the analysis by applying the RCM model to the QENS data. Comparing the model fitting and the experimental data, we see that the RCM analyses are satisfactory. The panels on the left-hand side and the right-hand side show the analysis results for two different temperatures under the same pressure, P = 400 bar. We can see clearly that for higher temperature case (T = 240K), the peak height is much lower than the lower temperature case (T = 210K). Since the structure factors are normalized before fitting, this fact implies that the peak wings of the high T spectra are broader than which of the low T spectra, thus there are more quasi-elastic components in the high T spectra. We can find that this is true in Fig. 2. Comparing the fitted results for the same Q value, the quasi-elastic components are clearly much broader in the high T cases.

From the results of RCM analysis of experimental $S_H(Q,\omega)$, we obtain three parameters, $\tau_0$, $\beta$, and $\gamma$, and are able to calculate the theoretical ISF from equation (1), which are plotted in Fig. 2, upper panels. They show clearly the Q-dependent two-step relaxation process (β-relaxation for the short time process and α-relaxation for the long time one) described by the RCM.

The temperature dependent behavior of $\langle \tau_T \rangle$ is shown in the upper six panels in Fig. 3 for six different pressures. For measurements under pressures below 1600 bar, at high temperatures, $\langle \tau_T \rangle$ obeys a Vogel-Fulcher-Tammann (VFT) law, $\langle\tau_T\rangle = \tau_0 \exp[DT_0/(T-T_0)]$, where D is a dimensionless parameter providing the measure of fragility and $T_0$ is the ideal glass transition temperature. For lower temperatures, the $\langle\tau_T\rangle$ switches over to an Arrhenius behavior, which is $\langle\tau_T\rangle = \tau_0 \exp(E_A/RT)$, where $E_A$, is the activation energy for the relaxation process and R is the gas constant. This dynamic crossover from the super-Arrhenius (the VFT law) to the Arrhenius behaviors defines the crossover temperature $T_L$ ($1/T_L = 1/T_0 - Dk_B/E_A$), much more sharply than that indicated by the MSD $\langle x_{H_2O}^2 \rangle$ vs. T plots shown in Fig. 1.

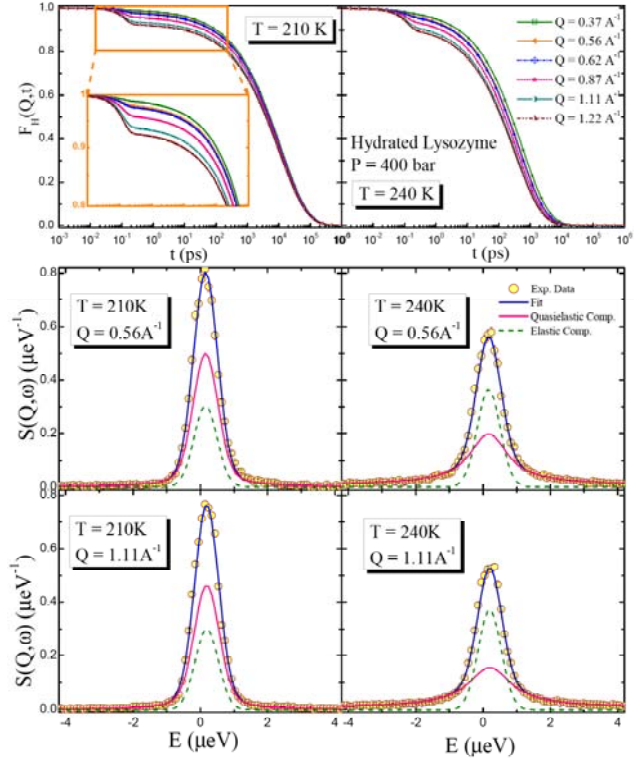

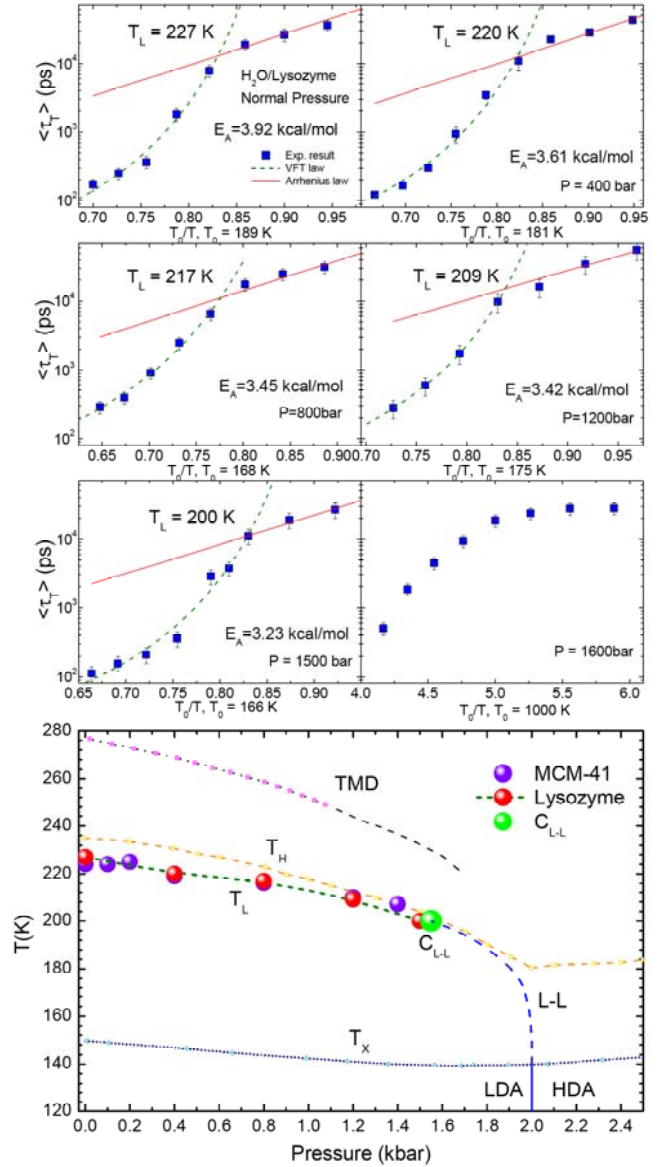

Fig. 2. The *lower four panels* show the RCM analyses of the normalized difference Spectra (contribution from hydration water). The *upper two panels* show the ISF of hydrogen atoms in the hydration water at different Q values under the pressure of 400 bar. We show the results at two temperatures: T = 210K (below $T_L$, left panels) and T = 240K (above $T_L$, right panels).

We have previously shown by a molecular dynamics (MD) simulation [16] that this super-Arrhenius to Arrhenius crossover is due to crossing of the Widom line in the one phase region. Upon the crossing of the Widom line, the local structure of water evolves from a predominately high density form (HDL, fragile liquid) to a predominately low density form (LDL, strong liquid) as the temperature crosses this characteristic temperature $T_L$ [23]. At the pressure of 1600 bar, $\langle\tau_T\rangle$ appears to be a smooth curve, neither having the super-Arrhenius behavior at high temperature nor Arrhenius behavior at low temperature. We may attribute it [14] to the phase separated mixture of the HDL and LDL due to the crossing of the hypothetical liquid-liquid first order transition line [24]. If these arguments are valid, then the disappearance of the FSC phenomena signals the crossing from the Widom line to the first order liquid-liquid transition line. These two lines are separated by the liquid-liquid critical point if it exists.

Fig. 3. *Upper six panels* show the extracted $\langle\tau_T\rangle$ plotted in a log scale against $T_0/T$ under six different pressures P = ambient pressure, 400, 800, 1200, 1500 and 1600bar. The *bottom panel* shows the pressure dependence of the measured FSC temperature $T_L$ (red circles), plotted in the T-P plane, comparing with our previous results from water in MCM-41-S [14](purple circles). We also show the homogeneous nucleation temperature line ($T_H$, [25]), crystallization temperatures of amorphous solid water ($T_X$, [26]), and the temperature of maximum density line (TMD, [27]), taken from known phase diagram of bulk water.

In the lower panel of Fig. 3, we thus plot the trajectory of the crossover temperature $T_L$ as a function of P (red circles). It is remarkable to see that this Widom line of the protein hydration water seems to coincide with the Widom line of the confined water in MCM-41-S found in our previous experiment [14].

In Fig. 4, we plot the value of $<\tau_T>$ at the crossover temperature $T_L$ as a function of pressure. It is surprising to find that although $T_L$ is pressure dependent, the $<\tau_T>$ values at $T_L$ are not changing with pressure. For all the pressures we measure, this value keeps constant at around 12 ns. We also plot the activation energy $E_A$, which is calculated from the Arrhenius behavior of $\langle\tau_T\rangle$ at low temperatures, and find it decreases with the pressure, which is coincident with previous MD simulation result [28]. Furthermore, we plot $E_A(P)/(k_BT_L)$ as a function of P. It appears to be independent of P again, within the error bars. The MD simulation [28] may be able to substantiate and elucidate its physical origin of this constancy.

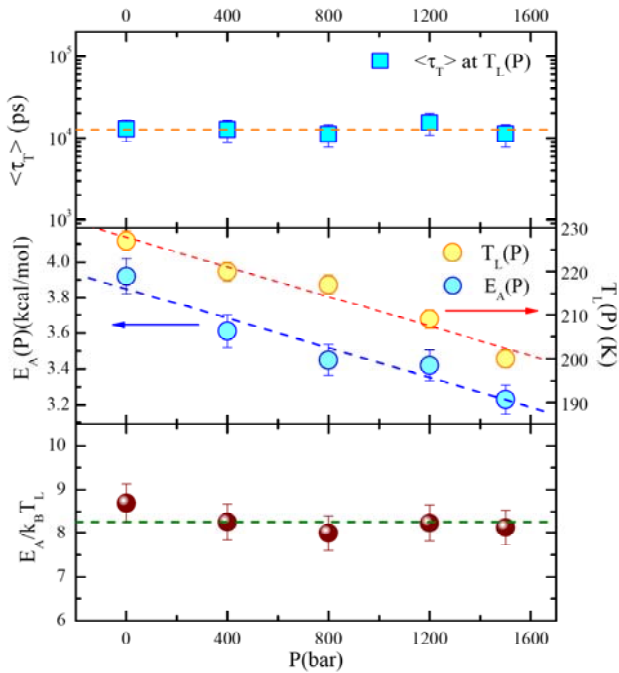

Fig. 4. Pressure Dependence of the parameters extracted from analysis of the quasi-elastic spectra by using Relaxing-Cage Model. Upper panel shows the pressure dependence of $\langle\tau_T\rangle$ at the corresponding crossover temperature $T_L(P)$. Middle panel shows the activation energy $E_A(P)$ and $T_L(P)$. Lower panel shows the ratio of the activation energy $E_A(P)$ to $k_BT_L$. The dashed lines indicate the linear least square fitting of these parameters.

In summary, we have made an extensive series of studies on the pressure dependence of the average translational relaxation time (or α-relaxation time) of protein hydration water by high-resolution QENS. We find strong evidence that the pressure dependence of the dynamic crossover temperature in protein follows that of the pressure dependence of the FSC temperature of hydration water. We find the Widom line of the 2-D confined protein hydration water is nearly coincident with the 1-D confined water in MCM-41-S with cylindrical pore of diameter 15Å. This strongly suggests that the liquid-liquid critical point of water $C_{L-L}$, if it exists, is located at approximately $T_C$ = 200 ± 5K, $P_C$ = 1550 ± 50 bars in both confined geometries. We also observe an abnormal deviation from the VFT law, just above the crossover temperature in the Arrhenius plot of $\langle\tau_T\rangle$ at P = 1500bar (see Fig. 3), which is very close to the predicted critical pressure. We may consider this deviation to come from the large density fluctuation of hydration water near the critical point. This may be another evidence that our predicted location of critical point is close to the actual one.

The research at MIT is supported by DOE Grants No. DE-FG02-90ER45429. This work utilized facilities supported in part by the National Science Foundation under Agreement No. DMR-0086210. We appreciate technical supports during experiments from V. Garcia-Sakai, T. Jenkins, M. Tyagi, and S. Poulton of NIST Center for Neutron Research. EF and PB acknowledge financial support from CSGI and MIUR. We benefited from affiliation with EU funded Mari-Curie Research and Training Network on Arrested Matter.

*Author to whom correspondence should be addressed. Email :sowhsin@mit.edu

**References:**
[1] B. F. Rasmussen, A. M. Stock, D. Ringe, and G. A. Petsko, Nature **357**, 423-424 (1992).
[2] R. M. Daniel, J. C. Smith, M. Ferrand, S. Hery, R. Dunn, and J. L. Finney, *Biophys. J.* **75**, 2504-2507 (1998).
[3] J. Pieper, T. Hauss, A. Buchsteiner, et al, *Biochemistry* **46**, 11398-11409 (2007).
[4] J. H. Roh, J.E. Curtis, S. Azzam, V. N. Novikov, I. Peral, Z. Chowdhuri, R. B. Gregory, and A. P. Sokolov, *Biophys. J.* **91**, 2573-2588 (2006).
[5] M. M. Teeter, *Annu. Rev. Biophys. Biophys. Chem.* **20**, 577, (1991).
[6] A. Paciaroni, A. R. Bizzarri, and S. Cannistraro, *Phys. Rev. E* **60**, R2476-R2479 (1999).
[7] G. Caliskan, A. Kisliuk, and A. P. Sokolov, *J. Non-Crys. Sol.* **307-310**, 868-873 (2002).
[8] P. W. Fenimore, H. Frauenfelder, et al, PNAS **101**, 14409-14413 (2004).
[9] P. W. Fenimore, et al, Physica (Amsterdam), **351A,** 1(2005).
[10] S. -H. Chen, L. Liu, E. Fratini, P. Baglioni, A. Faraone, and E. Mamontov, *PNAS USA* **103**, 9012-9016 (2006).
[11] S. -H. Chen, L. Liu, X. Chu, Y. Zhang, E. Fratini, P. Baglioni, A. Faraone, and E. Mamontocv, *J. Chem. Phys.* **125**, 171103-171106 (2006).
[12] X.-Q. Chu, E. Fratini, P. Baglioni, A. Faraone and S. -H. Chen, *Phys. Rev. E* **77**, 011908 (2008).
[13] K. Ito, C. T. Moynihan and C. A. Angell, Nature (London) **398**, 492(1999).
[14] L. Liu, S. -H. Chen, A. Faraone, C. -W. Yen, and C. -Y. Mou, *Phys. Rev. Lett.* **95**, 117802-117802 (2005).
[15] A. Faraone, L. Liu, C.-Y. Mou, C.-W. Yen, and S.-H. Chen, *J. Chem. Phys.* **121**, 10843 (2004).
[16] L. Xu, P. Kumar, S.V. Buldyrev, S.-H. Chen, P. H. Poole, F. Sciortino, and H. E. Stanley, *Proc. Natl. Acad. Sci.U.S.A.,* **102**, 16558-16562 (2005).
[17] G. Careri, *Prog. Biophys. Mol. Biol.* **70,** 223-249 (1998).
[18] M. Settles and W. Doster, *Faraday Discussion* **103**, 269-280 (1996).
[19] S. Dellerue and M.-C. Bellissent-Funel, Chemical Physics **258**, 315-325 (2000).
[20] A. Meyer, R. M. Dimeo, P. M. Gehring, and D. A. Neumann, *Rev. Sci. Instrum.* **74**, 2759-2777 (2003).
[21] K. Heremans, P. T. T. Wong, Chemical Physics Letters **118(1),** 101-4 (1985).
[22] S.-H. Chen, C. Liao, F. Sciortino, P. Gallo and P. Tartaglia, *Phys. Rev. E* **59**, 6708 (1999).
[23] F. Mallamace, M. Broccio, et al, *PNAS* **104,** 428 (2007).
[24] P. H. Poole *et al.*, *Nature (London)* **360**, 324 (1992).
[25] H. Kanno *et al.*, Science **189**, 880 (1975).
[26] H. E. Stanley, *Mysteries of Water*, M.-C. Bellissent-Funel (Ed.), Nato Science Series A **305** (1999).
[27] C. A. Angell *et al.*, J. Non-Cryst. Solids **205-207**, 463 (1996).
[28] P. Kumar, G. Franzese and H. E. Stanley, *Phys. Rev. Lett.* **100**, 105701 (2008)).